\documentclass[twocolumn,showpacs,aps,prb]{revtex4}
\usepackage{txfonts}
\usepackage{graphicx}
\usepackage{dcolumn}
\usepackage{bm}
\usepackage{color}
\usepackage{marvosym}

\begin{document}
\preprint{Physical Review B} %

\title{Correlations and entanglements in a few-electron quantum dot without Zeeman splitting}

\author{Ning Yang}
\affiliation{%
Department of Physics, Key Lab of Atomic and Molecular Nanoscience,
and Center for Quantum Information, Tsinghua University, Beijing
100084, People's Republic of China}
\author{Jia-Lin Zhu}
\email[Electronic address: ]{zjl-dmp@tsinghua.edu.cn}
\affiliation{%
Department of Physics, Key Lab of Atomic and Molecular Nanoscience,
and Center for Quantum Information, Tsinghua University, Beijing
100084, People's Republic of China}
\author{Zhengsheng Dai}
\affiliation{%
Department of Physics, Key Lab of Atomic and Molecular Nanoscience,
and Center for Quantum Information, Tsinghua University, Beijing
100084, People's Republic of China}
\author{Yuquan Wang}
\affiliation{%
Department of Physics, Key Lab of Atomic and Molecular Nanoscience,
and Center for Quantum Information, Tsinghua University, Beijing
100084, People's Republic of China}

\date{\today}

\begin{abstract}
We explore the correlations and entanglements of exact-diagonalized
few-electron wave functions in a quantum dot in magnetic fields
without the Zeeman splitting. With the increase of the field, the
lowest states with different spins gradually form a narrow band and
the electronic states undergo a transition from liquids to rotating
Wigner molecules which are accompanied by different characters of
charge correlations. For both the liquid and crystal states, the
spin conditional probability densities show magnetic couplings
between the particles which depend on the particle numbers, the
total spins and the angular momenta of the states. The von Neumann
entropies show the spin-dependent entanglements between electrons.
The regular magnetic-coupling oscillations and converging
entanglement entropies emerge in the rotating Wigner molecular
states.
\end{abstract}
\maketitle
\section{Introduction}
Driven by the interests in both basic research and technological
application, the electronic structures of two-dimensional quantum
dots (QDs) have been a topic extensively studied in recent years.
The particle number in QDs can be reduced precisely down to a few
electrons with highly controllable confinements, interactions and
external fields. The specific spin states of electrons and their
manipulations by magnetic and electric fields have been proposed for
basic qubit schemes\cite{Loss1998, Kyriakidis2005} for future
quantum computation. Therefore, it is important to understand the
quantum behaviors of few-electron QDs. Theoretically, the system
with appropriate magnetic fields may be viewed as a finite-size
precursor of integer and fractional quantum Hall states in
two-dimensional electron gas (2DEGs) which can be understood by the
laughlin wave functions\cite{Laughlin1983} and composite fermion
theory.\cite{Jain1989} In the ordinary fractional Hall system, the
short-range interactions between particles are important. However,
in few-electron QDs, long-range interactions are important due to
the low charge density, and make the characters of confined
electrons different from those of extent system with larger charge
density. In magnetic fields, it has been demonstrated that there are
the transitions from liquid states to rotating Wigner molecular
(RWM) states\cite{Yannouleas2002, Tavernier2003, Reimann2006} in
QDs. Similar liquid-crystal transitions are also expected for 2DEGs
but in very strong fields.\cite{Yang2001, Mandal2003, He2005}

Both the fractional quantum Hall states and the electronic states in
QDs are strongly correlated. Here, the correlations mean not only
the classic but also the quantum ones, i.e. the entanglements
between electrons. The essential role of the quantum correlation in
fractional quantum Hall system has been widely discussed. However,
the investigations of the entanglements in few-electron QDs are
still limited.

When discussing QDs in strong magnetic fields, the ground states can
be often taken as full polarized due to the Zeeman splitting. With
improving nanotechnology, it is achieved to fabricate the QDs with
negligible Zeeman splitting in recent experiments.\cite{Salis2001,
Ellenberger2006} Then the electrons of the ground states in strong
magnetic fields may not be full polarized. With the spin degree of
freedom, although the liquid-crystal transitions still exist, both
the characters of energy level structure and correlations between
electrons are different from the full polarized systems. In this
work, based on exact diagonalization (ED) method, we explore the
correlations and entanglements in the QDs with spin degree of
freedom and discuss their behaviors in the liquid-crystal
transition.

\section{Model and Formula}
The Halmiltonian of a N-electron QD in a perpendicular magnetic
field without the Zeeman splitting is written as
\begin{equation}\label{eq01}
H\!\!=\!\!\sum_{i=1}^N {\left [\frac{1}{2m}\left
(\hat{P}_i+e\vec{A}\right )^2\!\!+V(r_i)\right
]}\!\!+\!\!\sum_{i<j}{\frac{e^2}{4\pi\varepsilon|\vec{r}_i-\vec{r}_j|}}.
\end{equation}
The first and second parts of Eq.(\ref{eq01}) are respectively the
single-particle energies and interaction energies of the electrons.
$\vec{A}$ is the vector potential of the magnetic field. $m$ and
$\epsilon$ are the effective mass and static dielectric constant
which are respectively 0.067$m_e$ and 12.4 for GaAs. $V$ is the
confinement of the dot.

We use the method of series expansion\cite{Zhu1997} to get exact
eigenstates $\varphi_\alpha(\vec{r})$ of the single-particles parts
of Eq.(\ref{eq01}) as a set of single-particle bases. Then in the
second quantization scheme the Hamiltonian(\ref{eq01}) can be
written as
\begin{equation}\label{eq02}
H=\sum_{\alpha} {\epsilon_\alpha a_\alpha^\dag a_\alpha} +
\frac{1}{2} \sum_{\alpha \beta \alpha' \beta'}
{g_{\alpha'\beta'\beta\alpha}a_{\alpha'}^\dag a_{\beta'}^\dag
a_\beta a_\alpha}
\end{equation}
with
\begin{displaymath}
g_{\alpha'\beta'\beta\alpha}=\int{d\vec{r_1}d\vec{r_2}\varphi_{\alpha'}^*(\vec{r_1})
\varphi_{\beta'}^*(\vec{r_2}) \frac{e^2}{4\pi \epsilon
|\vec{r_1}-\vec{r_2}|} \varphi_\beta(\vec{r_2})
\varphi_\alpha(\vec{r_1})}
\end{displaymath}
where $a_\alpha^\dag(a_\alpha)$ is the creation (annihilation)
operator of the single-particle state $\varphi_\alpha$, and
$\epsilon_\alpha$ is its energy. Then Eq.(\ref{eq02}) is
diagonalized to obtain the energies and the corresponding wave
functions $\Psi$ of the few-electron states. In the following
discussions, the confinement of the dot is parabolic and the
strength is 2 meV. It is worthwhile to point out that the series
expansion method is applicable not only to the parabolic
confinement, but also other confinement forms for which analytic
single-particle bases cannot be obtained. Without spin-orbit
coupling, the few-electron states can be the common eigenstates of
total angular momentum $L$, total spin $S$ and its z-component
$S_z$. In the paper, we will mark the lowest states with different
spins as ($L$,$S$,$S_z$) for brevity. Of course, the states with
same $L$ and $S$ but different $S_z$ are degenerate due to the
absence of the Zeeman splitting.

Having got the eigenstates of Eq.(\ref{eq02}), we can evaluate the
charge and spin correlations by the conditional probability
densities (CPDs)
\begin{equation}
P(r,r')=\frac{1}{2}\langle\Psi|\sum_{\alpha \beta \alpha' \beta'}
\delta(\vec{r}-\vec{r_i})\delta(\vec{r'}-\vec{r_j})a_{\alpha'}^\dag
a_{\beta'}^\dag a_\beta a_\alpha|\Psi\rangle
\end{equation}
and
\begin{eqnarray}
\lefteqn{P(r,r';\sigma,\sigma')=}\nonumber\\
&&\frac{1}{2}\langle\Psi|\!\!\sum_{\alpha \beta \alpha' \beta'}\!\!
\delta(\vec{r}-\vec{r_i})\delta_{\sigma\sigma_i}\delta(\vec{r'}-\vec{r_j})
\delta_{\sigma'\sigma_j}a_{\alpha'}^\dag a_{\beta'}^\dag a_\beta
a_\alpha|\Psi\rangle.
\end{eqnarray}

In this paper, we also investigate the behaviors of entanglements of
the few-electron electronic states. It has been demonstrated that
the von Neumann entropy\cite{Paskauskas2001}
$S=-\text{tr}[\rho^f\ln\rho^f]$ can be used for quantifying the
entanglements between particles. $\rho^f$ is the single-particle
reduced density matrix. Although in multiparticle case it is not
convinced how to quantify all the entanglement properties of an
identical-particle state, the von Neumann entropy can still give the
entanglement information between one particle and the other part of
the system. In the following discussions, we employ a modified form
of the von Neumann entropy as\cite{Zeng2002}
\begin{equation}\label{entform}
S=-\text{tr}[\rho^f\ln\rho^f]-\ln N
\end{equation}
with
\begin{equation}
\rho_{\alpha\beta}^f=\langle\Psi|a_\alpha^\dag a_\beta|\Psi\rangle
\end{equation}
where N is the particle number of the system. With such
modification, the entropy due to the indistinguishability of the
particles is subtracted. And the lower limit values of entropies for
the system with different particle number are all equal to zero,
which corresponds to the unentangled states.

\section{Results and Discussion}
\subsection{Energy level structures}
\begin{figure}[ht]
\includegraphics*[angle=0,width=0.45\textwidth]{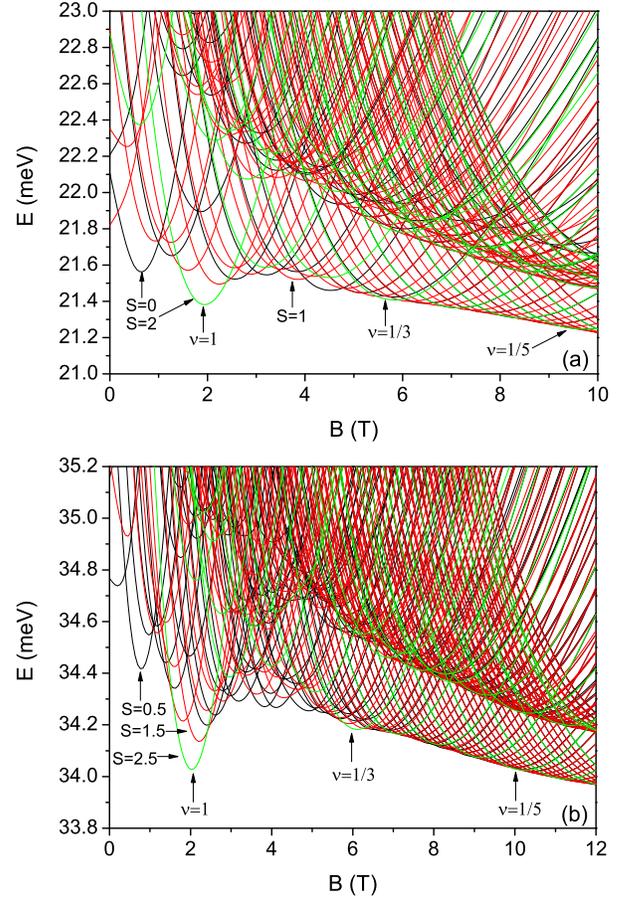}
\caption{\label{FIG:EnergySpectrum} (Color online) Energy spectra of
the four- (a) and five-electron (b) states as functions of the
magnetic field. N times of the energy of the first Landau level have
been subtracted from the total energy. Black, red and green lines
correspond to the four- (five-) electron states with S=0 (0.5), 1
(1.5) and 2 (2.5), respectively. The states correspond to the
filling factors $\nu=1$, 1/3 and 1/5 have been indicated.}
\end{figure}

The energy spectra of four- and five-electron QDs in magnetic fields
are shown in Fig.\ref{FIG:EnergySpectrum}. For clarity, N times of
the energy of the first Landau level have been subtracted from the
total energy. In smaller fields, the states with different spins are
clearly separated. The full-polarized ground state corresponding to
filling factor $\nu=1$, namely the maximum density droplet
(MDD)\cite{Oosterkamp1999, Tavernier2003}, has lower energy in a
long range of magnetic field. Without the Zeeman splitting, the
states with lower spin can become the ground states even in strong
magnetic fields. The ground states in the magnetic fields
corresponding to fractional filling factors (1/3, 1/5, etc.) are
still full-polarized. In the magnetic fields where the flux deviates
from the values of fractional filling factors by $\pm\phi_0$
($\phi_0$ is the quantum of magnetic flux) the ground states will be
the ones with lowest total spin. Such situation is same as that in
2DEGs with spin degree of freedom. In strong fields, the electrons
form the RWMs. With the decreased overlapping of electrons, the
exchange energies of different spin states are depressed totally.
Then the states with different total spins are almost degenerate and
form a narrow band.

\begin{figure*}[ht!]
\includegraphics*[angle=0,width=0.75\textwidth]{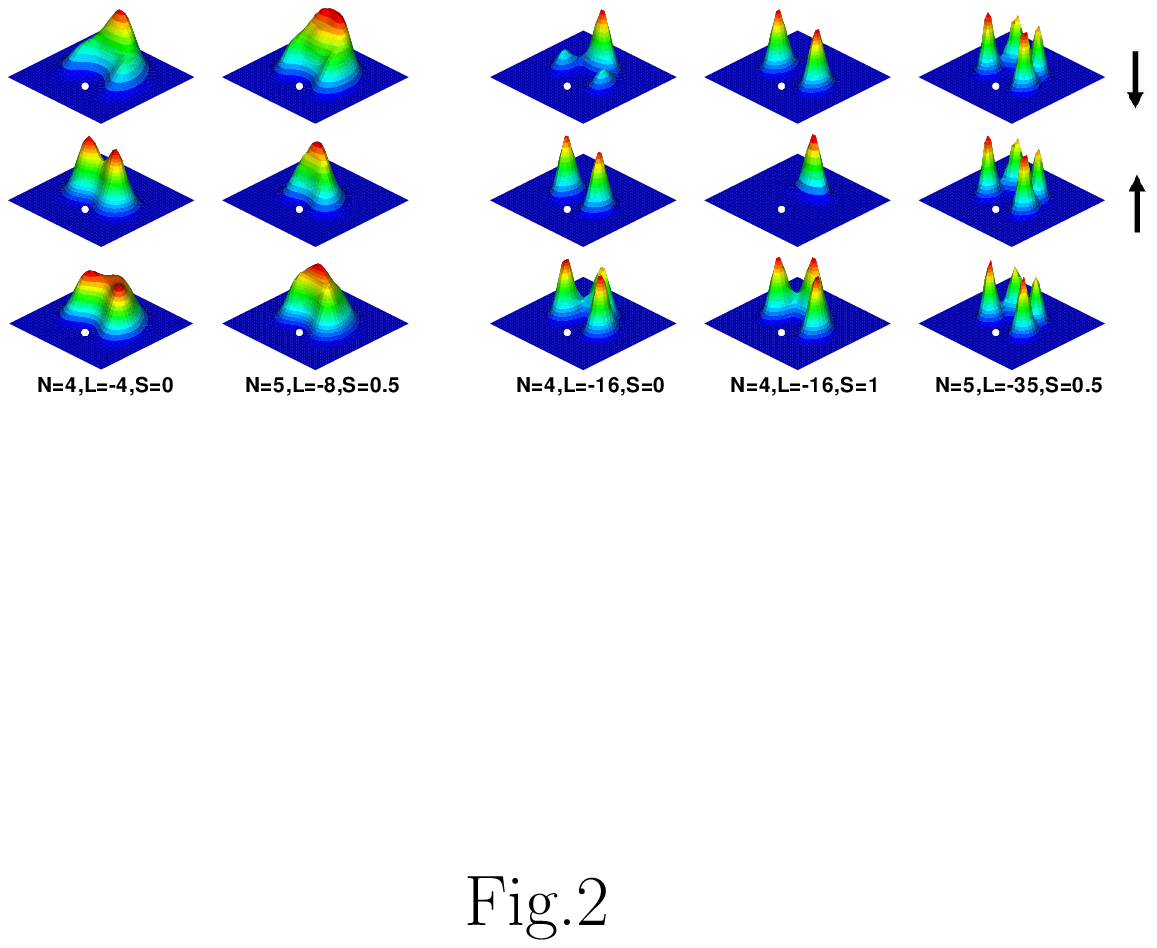}
\caption{\label{FIG:Pcf} (Color online) First two rows: spin
conditional probability densities of finding spin-down (first row),
spin-up (second row) electrons with a spin-up electron fixed at the
position indicated by the white dot. Third row: conditional
probability densities of finding electrons with an electron fixed at
the position indicated by the dot.}
\end{figure*}

With the increase of the field, there are angular momentum
transitions for the ground states. The allowable angular momenta are
so called magic numbers.\cite{Ruan1995} With the Zeeman splitting,
the angular momentum transition of the four- and five-electron
ground states in magnetic fields has the period as the particle
number N. Including the spin degree of freedom, the angular momentum
transition of ground states with the increase of the field is almost
continuous. However, for the lowest states with different total
spins, the angular momentum transitions indeed have some rules. In
strong magnetic fields where the electron are well localized, such
rules can be obtained from the theory of electron
molecules.\cite{Maksym1995, Maksym2000} For example, the lowest
four-electron states with $S=0$ must be the ones with angular
momenta $|L|=4n$ or $4n+2$, the lowest states with $S=1$ must be the
ones with $|L|=4n$ or $4n\pm1$. For five electrons, it can be
obtained from our studies that the angular momenta of lowest states
with $S=0.5$ can be the continuous integers, the lowest states with
$S=1.5$  must be the ones with $L\neq5n$, where $n$ is an integer.
In both cases, the transitions for full-polarized states still have
a period of particle number N.

In small magnetic fields where the electronic states are still
liquid-like ones, the transition rules of angular momenta described
above are not exact. The states with some angular momenta coincide
with the rules do not appear as the lowest states. For example, for
the five-electron case with $S=0.5$, the angular momenta can be
arbitrary integers in strong fields, but the states with
$|L|=2,3,6,9,10,11,14,17,18,20,22,23,25,29$ do not appear as the
lowest states. Examining the results obtained by exact
diagonalization carefully, it can be found that such deletion for
four- (five-) electron states disappears when the filling factor
$\nu<1/2\ (1/3)$. The different angular momentum transition patterns
just reflect the transition of the electronic states from the liquid
to crystal ones with the increase of the field. We also illustrate
the angular momentum transition pattern for the lowest states with
different spins in Fig.\ref{FIG:MagneticCoupling}. It can be seen in
both Fig.\ref{FIG:EnergySpectrum} and Fig.\ref{FIG:MagneticCoupling}
that the angular momentum transitions become regular in strong
fields, as discussed above.

\subsection{Spin correlations and magnetic couplings}
Within the liquid-crystal transition, the correlations between
electrons of the lowest states change from the short-range liquid to
long-range crystal ones. In the third row of Fig.\ref{FIG:Pcf}, we
show some examples of the two kinds of correlations. It can be seen
in the plot that the CPDs of the four-electron state (-4,0,0) and
five-electron state (-8,0.5,0.5) are liquid-like. In strong magnetic
fields, the states with larger angular momenta exhibit crystal-like
correlations, which are the characters of the RWMs.

Besides the angular momentum and charge correlation transitions,
there are also different spin correlations. We inspect the spin
correlations by examining the spin CPDs to find electrons with a
certain spin when a spin-up electron is fixed. We found that the
spin correlations can reveal the magnetic couplings between
electrons in non- and partial-polarized states with minimum $S_z$.
We illustrate some spin CPDs in the first two rows of
Fig.\ref{FIG:Pcf}. It can be seen that when a spin-up electron is
fixed at the position indicated by the white dot, the spin-up and
spin-down electrons in the state (-16,0,0) will be probably at the
neighbor and opposite positions, respectively. Such CPDs just reveal
the existence of ferromagnetic couplings between electrons. For the
state (-16,1,0), the CPDs shown in the figure are just contrary to
that for (-16,0,0) and reveal the anti-ferromagnetic couplings. For
the five-electron state (-35,0.5,0.5), the spin CPDs show that
neither the spin-up nor the spin-down electrons have preferable
positions when a spin-up electron is fixed. Then we cannot expect
any specific magnetic coupling for it.

We have found that other states also exhibit respective couplings
even if the angular momenta are so small that the states are still
liquid ones. As shown in Fig.\ref{FIG:Pcf}, the CPDs for the state
(-4,0,0) and (-8,0.5,0.5) also reveal the ferromagnetic couplings
between electrons. Although the electrons are no longer well
localized, the spin-up electrons are closer to the fixed electron
than the spin-down ones.

\begin{figure}[ht]
\includegraphics*[angle=0,width=0.43\textwidth]{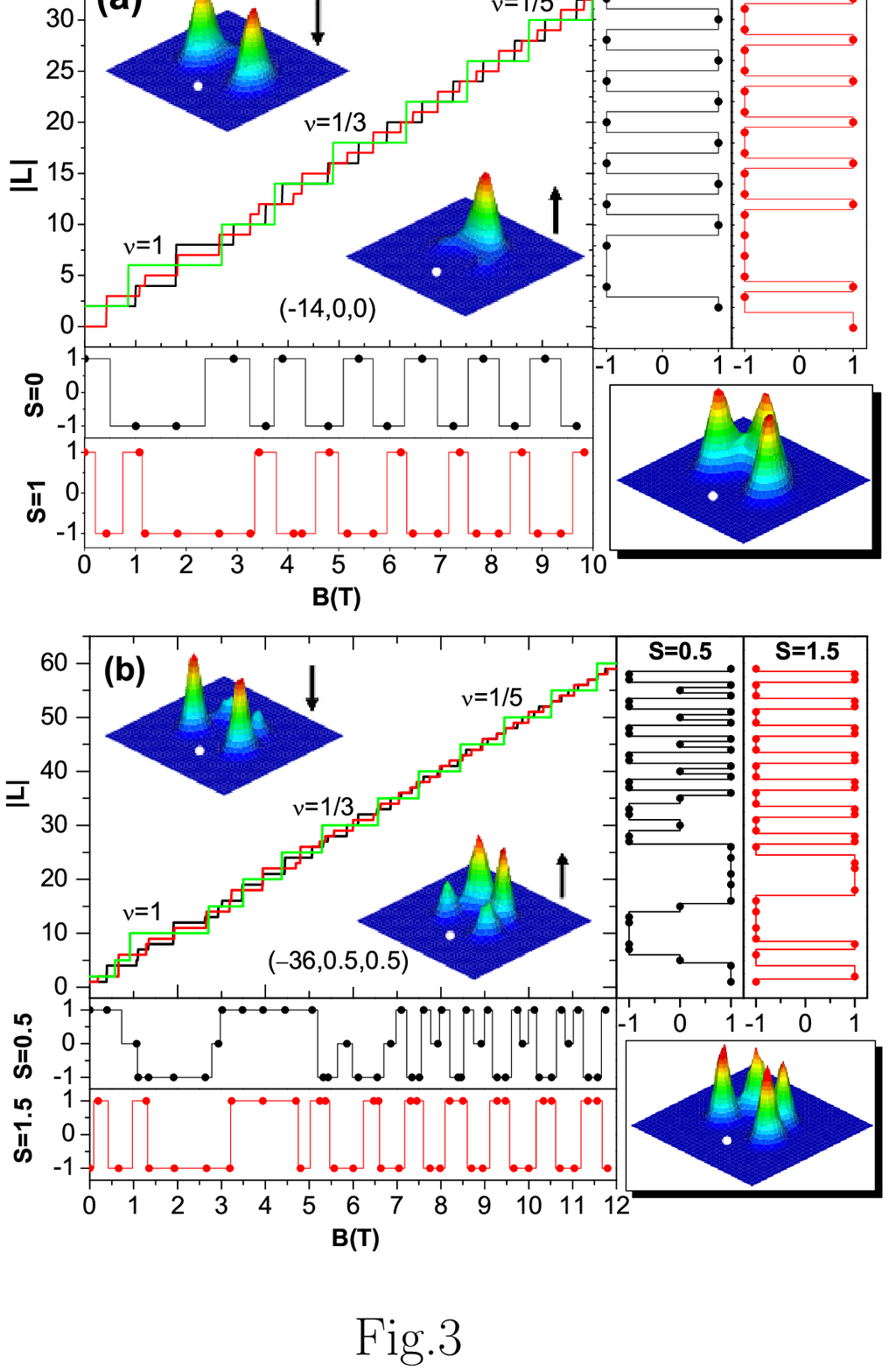}
\caption{\label{FIG:MagneticCoupling}(Color online) Angular momentum
transitions of four-electron (a) and five-electron (b) states as
functions of magnetic fields. Black, red and green lines correspond
to the states with $S=0, 1$ and 2 for four electrons and 0.5, 1.5
and 2.5 for five electrons. The positions correspond to the filling
factor $\nu=1$, 1/3 and 1/5 have been indicated. The magnetic
coupling oscillations of the non- and partial-polarized states with
minimum $S_z$ as function of magnetic field and angular momentum are
shown in the bottom and right panels, respectively. Insets: the spin
CPDs of the states (-14,0,0) and (-36,0.5,0.5) for spin-down and -up
electrons and the charge CPDs are shown from left-upper to
right-bottom corner.}
\end{figure}

It should be pointed out that, strictly speaking, the four-electron
states can perform the ferromagnetic and anti-ferromagnetic
couplings. For five-electron states, only ferrimagnetic coupling
along with the ferromagnetic coupling exist because of the unequal
electron numbers with two species of spins. One example of CPDs
showing the ferrimagnetic coupling are presented in the inset of
Fig.\ref{FIG:MagneticCoupling}(b). And full-polarized states cannot
exhibit any specific magnetic coupling.

In the previous discussions, we have shown that the spin CPDs can
reveal the magnetic couplings in the lowest states with different
spins although the correlations are quite different for the liquid
and crystal ones. Then we can identify the specific coupling for
each state.  The magnetic couplings as functions of magnetic fields
i.e. angular momenta are shown in the bottom and right panels of
Fig.\ref{FIG:MagneticCoupling}. The values $1, -1,\text{and } 0$
correspond to the anti-ferromagnetic (ferrimagnetic), ferromagnetic
and no specific coupling, respectively. For the states with
different angular momenta, the couplings are different. And it is
interesting to observe various magnetic coupling oscillation
patterns for the states with different total spins. Again, all
oscillations become regular with respect to the magnetic field, i.e.
angular momentum when the field is strong enough to form the RWMs.
In strong fields, the four-electron states with $S=0$ and $|L|=4n\
(4n+2)$ are ferromagnetic (anti-ferromagnetic).\cite{Maksym2000} The
states with $S=1$ and $|L|=4n\ (4n\pm1)$ are anti-ferromagnetic
(ferromagnetic), where $n$ is an arbitrary integer. The
five-electron states also have their own magnetic coupling rules as
shown in Fig.\ref{FIG:MagneticCoupling}. The states with $S=0.5$ and
$|L|=5n\pm1\ (5n\pm2)$ are ferrimagnetic (ferromagnetic). The
couplings for the states with $S=1.5$ are just contrary to that with
$S=0.5$. In any case, the oscillation with respect to the momenta
performs a period of particle number $N$. A special feature of the
five-electron case is that the states with $S=0.5$ and $|L|=5n$ have
no specific magnetic coupling, as the example (-35,0.5,0.5) in
Fig.\ref{FIG:Pcf}. Such systematic vanish of the magnetic coupling
is absent in the four-electron case.

In small fields, the oscillations are no longer regular. This is
because that the angular momentum transition pattern in small fields
is different from that in strong fields. The vanishing states with
some angular momenta cause the irregularity of the oscillation.

\subsection{Entanglement}
\begin{figure}[ht]
\includegraphics*[angle=0,width=0.38\textwidth]{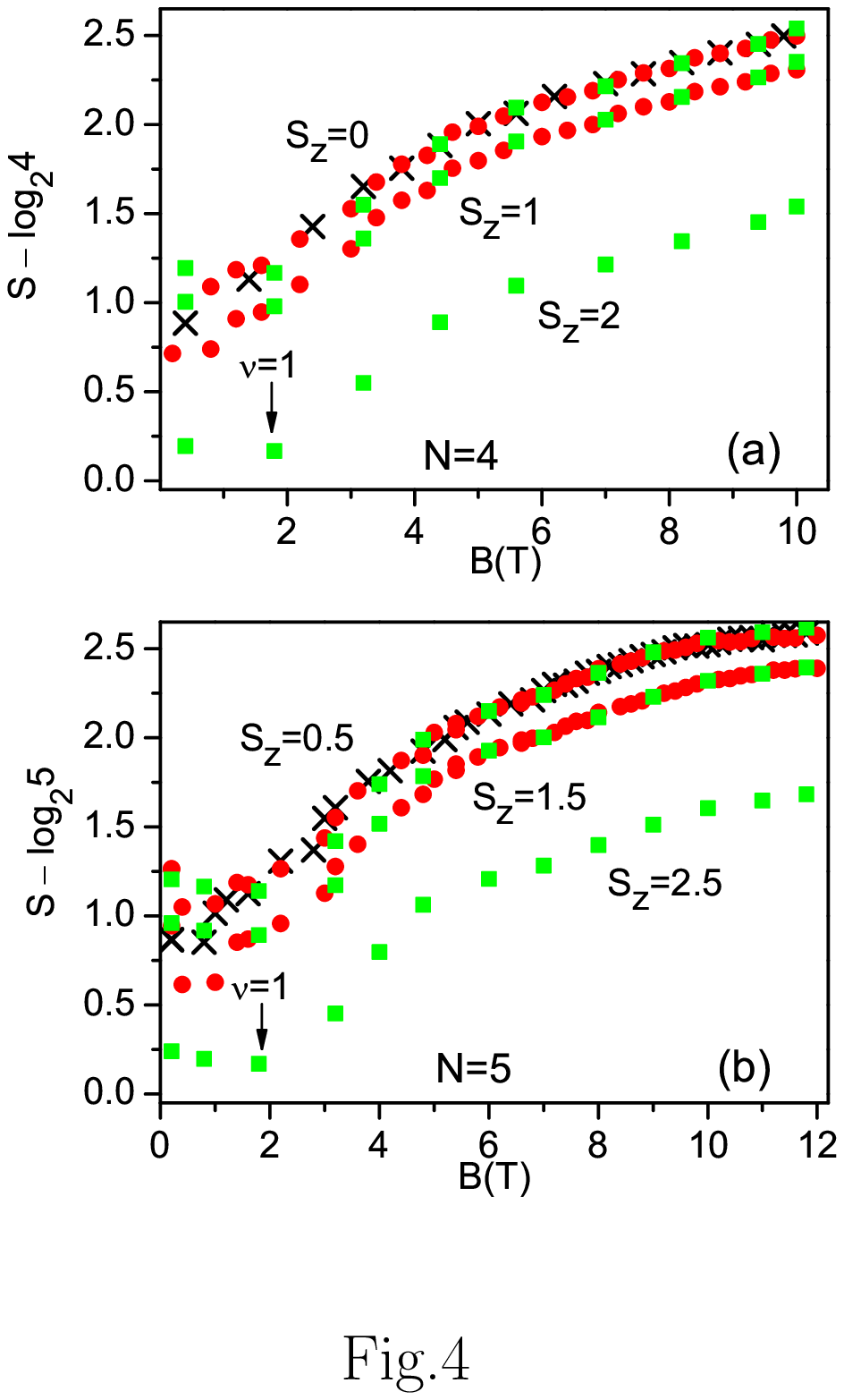}
\caption{\label{FIG:Entanglement} (Color online) Entanglement
entropies of the lowest states of four (a) and five (b) electrons
with different spins. A term $\log_2N$ has been subtracted.
$\times$, $\bullet$ and {\Rectsteel} correspond to the states with
non-, partial to full-polarized spins, respectively.}
\end{figure}
In this subsection, we investigate the features of entanglements in
the liquid-crystal transition. The von Neumann
entropy\cite{Paskauskas2001} is used for the quantification of the
extent of the entanglement. According to Eq.(\ref{entform}), the
entanglement entropies of different spin states as functions of the
magnetic field are given in Fig.\ref{FIG:Entanglement}.

A feature of the entanglements in QD is that the entropies of the
states with same $L$ and $S$ but different $S_z$ are unequal
although their energies are same when the Zeeman splitting is
ignored. The differences corresponding to different $S_z$ are
approximately constant in strong magnetic fields. Especially for a
state with maximum $S_z$ where the entropy owing to the spin
components is totally eliminated, its entropy should differ from
that of the state with minimum $S_z$ by one, which is revealed by
our calculation. Without the Zeeman splitting, the energies of the
states with same $L$ but different $S$ in strong magnetic fields are
nearly degenerate. In Fig.\ref{FIG:Entanglement}, the entropies of
such states with same $S_z$ also converge and increase monotonously.
Such convergence of the entanglements is also a character of RWMs.

For the liquid states in lower magnetic fields where the interaction
energies vary in different spin states, the variations of
entanglements are not monotonous any more. The states with different
$S$ have their own minimal values in respective fields. The state
corresponding to the MDD with maximum $S_z$ has the global minimum
entanglement entropy, i.e. the minimum correlation. In 2DEGs, the
Laughlin wave function with filling factor $\nu=1$ is
unentangled.\cite{Zeng2002} Similarly, in few-electron QDs, the
entanglement entropies of the MDD states are also very close to
zero.

\section{Summary}
To conclude, we have investigated the liquid-crystal transitions in
the few-electron quantum dot without the Zeeman splitting. The spin
degree of freedom brings various characters to the transition. The
energy level structures in the range of liquid and crystal phase are
quite different. In strong magnetic fields, different spin states
with specific angular momentum transition rules form a narrow band
which is the character of the rotating Wigner molecular states. In
small fields, such rules are no longer strictly obeyed by the
liquid-like states. For both the liquid- and crystal-like states
with the lowest $S_z$, there are magnetic couplings between
electrons although the spin CPDs are quite different. The species of
the couplings depend on the particle numbers, the total spins and
the angular momenta of the states. In the RWMs, the magnetic
couplings oscillate regularly with respect to the field. The
entanglement entropies which do not depend on the total spins of the
states increase monotonously. The entropy differences due to the
different $S_z$ are approximately constant. In the liquid states,
the oscillations of the magnetic couplings are irregular and the
variations of the entanglements are not monotonous. The studies
imply that the extent works on QDs with different Zeeman splittings
and other spin-related terms will be important for understanding and
controlling the quantum states of few electrons in the future.

\begin{acknowledgments}
Financial supports from NSF China (Grant No. 10574077), the``863"
Programme of China (No. 2006AA03Z0404) and the ``973" Programme of
China (No. 2005CB623606) are gratefully acknowledged.
\end{acknowledgments}

\appendix*



\begin{thebibliography}{}
\bibitem{Loss1998} D.Loss and D. P. DiVincenzo, Phys. Rev. A {\bf
57}, 120 (1998).

\bibitem{Kyriakidis2005} J. Kyriakidis and S. J. Penney, Phys. Rev. B {\bf
71}, 125332 (2005).

\bibitem{Laughlin1983} R. B. Laughlin, Phys. Rev. Lett. {\bf
50}, 1395 (1983).

\bibitem{Jain1989} J. K. Jain, Phys. Rev. Lett. {\bf
63}, 199 (1989).

\bibitem{Yannouleas2002} C. Yannouleas and U. Landman, Phys. Rev. B {\bf
66}, 115315 (2002).

\bibitem{Tavernier2003} M. B. Tavernier, E. Anisimovas, F. M. Peeters, B. Szafran, J. Adamowski, and S.
Bednarek, Phys. Rev. B {\bf 68}, 205305 (2003).

\bibitem{Reimann2006} S. M. Reimann \emph{et al.}, New J. Phys. {\bf
8}, 59 (2006).

\bibitem{Yang2001} K. Yang, F. D. M. Haldane, and E. H. Rezayi,
Phys. Rev. B {\bf 64}, 081301 (2001).

\bibitem{Mandal2003} S. S. Mandal, M. R. Peterson, and J. K. Jain,
Phys. Rev. Lett. {\bf 90}, 106403 (2003)

\bibitem{He2005} W. J. He, T. Cui, Y. M. Ma, C. B. Chen, Z. M. Liu, and G. T.
Zou, Phys. Rev. B {\bf 72}, 195306 (2005)

\bibitem{Salis2001} G. Salis \emph{et al.}, Nature. {\bf 414}, 619 (2001).

\bibitem{Ellenberger2006} C. Ellenberger \emph{et al.}, Phys. Rev. Lett. {\bf
96}, 126806 (2006).

\bibitem{Zhu1997} J. -L. Zhu, Z. -Q. Li, J. -Z. Yu, K. Ohno, and Y.
Kawazoe, Phys. Rev. B {\bf 55}, 15819 (1997).

\bibitem{Paskauskas2001} R. Pa\v{s}kauskas and L. You, Phys. Rev. A {\bf
64}, 042310 (2001).

\bibitem{Zeng2002} B. Zeng, H. Zhai, and Z. Xu, Phys. Rev. B {\bf
66}, 042324 (2002).

\bibitem{Oosterkamp1999} T. H. Oosterkamp \emph{et al.}, Phys. Rev. Lett. {\bf
82}, 2931 (1999).

\bibitem{Ruan1995} W. Y. Ruan, Y. Y. Liu, C. G. Bao, and Z. Q.
Zhang, Phys. Rev. B {\bf 51}, 7942 (1995).

\bibitem{Maksym1995} P. A. Maksym, Phys. Rev. B {\bf 53}, 10871
(1995).

\bibitem{Maksym2000} P. A. Maksym, H. Imamura, G. P. Mallon, and H.
Aoki, J. Phys.:Condens. Matter {\bf 12}, R299 (2000).

\end{thebibliography}
\end{document}